\documentclass{camera}
\usepackage{graphicx}  
\newcommand{\Msun}{$\mathrm{M}_{\odot}$}
\begin{document}

\title{Probing the equation of state of neutron-rich matter with
intermediate energy heavy-ion collisions}
\author{\small{Bao-An Li$^{1}$, Lie-Wen Chen$^{2}$, Che Ming Ko$^{3}$
\and Andrew W. Steiner$^{4}$}}

\organization{$^{1}$ Department of Chemistry and Physics, P.O. Box 419,\\
Arkansas State University, State University, AR 72467-0419, USA\\
$^{2}$ Institute of Theoretical Physics, Shanghai Jiao Tong University,
Shanghai 200240, and Center of Theoretical Nuclear Physics, National
Laboratory of Heavy-Ion Accelerator, Lanzhou, 730000, China\\
$^{3}$ Cyclotron Institute and Physics Department, Texas A\&M University,
College Station, Texas 77843, USA\\
$^{4}$ Theoretical Division, Los Alamos National Laboratory, \\
Los Alamos, NM 87545, USA}
\maketitle
\begin{abstract}
Nuclear reactions induced by stable and/or radioactive neutron-rich 
nuclei provide the opportunity to pin down the equation of state  
of neutron-rich matter, especially the density ($\rho$) dependence of 
its isospin-dependent part, i.e., the nuclear symmetry energy
$E_{\rm sym}$. A conservative constraint, $32(\rho /\rho_{0})^{0.7} 
< E_{\rm sym}(\rho ) < 32(\rho /\rho _{0})^{1.1}$, around the nuclear 
matter saturation density $\rho_0$ has recently been obtained from the 
isospin diffusion data in intermediate energy heavy-ion collisions. 
We review this exciting result and discuss its consequences and 
implications on nuclear effective interactions, radii and cooling 
mechanisms of neutron stars.
\end{abstract}

\section{Nuclear equation of state and symmetry energy of 
neutron-rich matter}\label{intro}

\begin{figure}[htb]
\includegraphics[width=6cm,height=5.5cm]{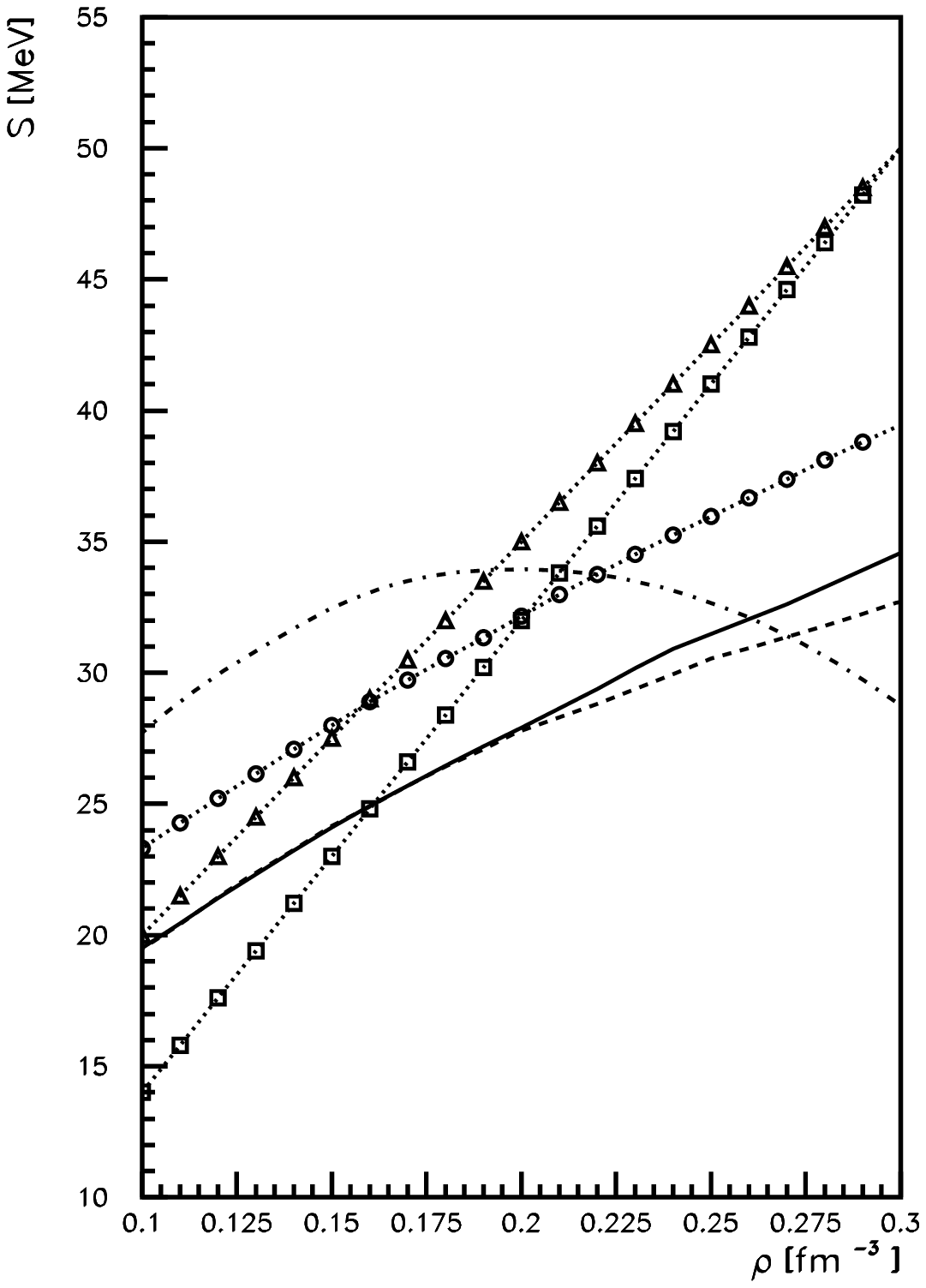}
\includegraphics[width=6cm,height=5.1cm]{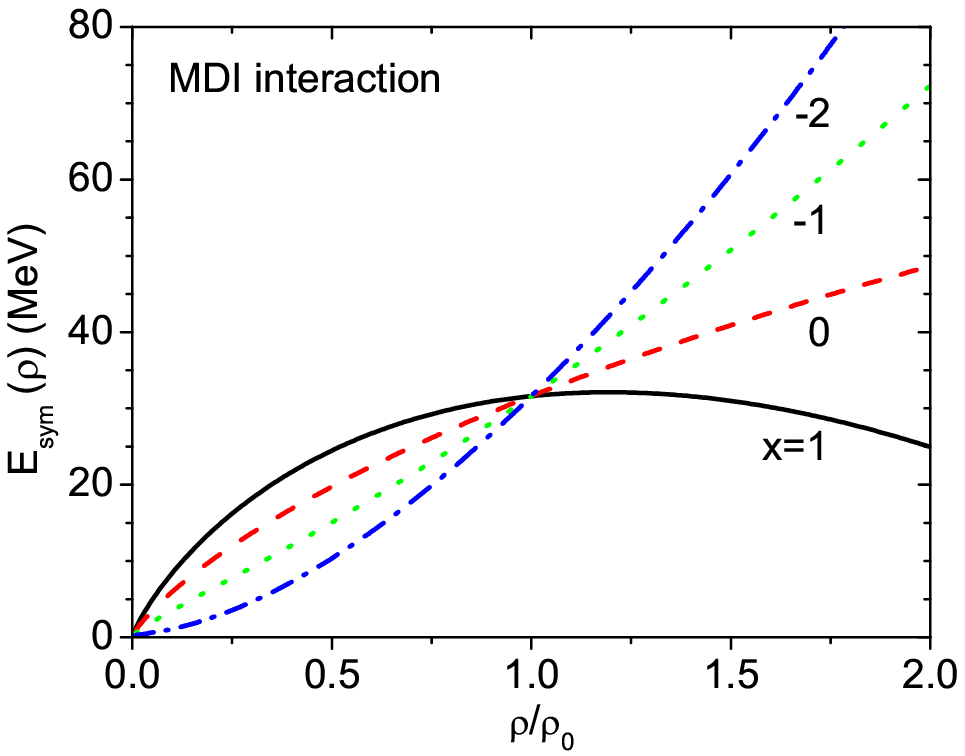}
\caption{{\protect\small Left panel: density dependence of the 
symmetry energy $S$ from the continuous choice Brueckner-Hartree-Fock 
with Reid93 potential (circles), self-consistent Green's function 
theory with Reid93 potential (full line), variational calculation 
with Argonne Av14 potential (dashed line), Dirac-Brueckner-Hartree-Fock 
calculation (triangles), relativistic mean-field model (squares), and
effective field theory (dash-dotted line)~\cite{diep}. Right
panel: same as left panel for the MDI interaction with four $x$ 
parameters as used in the IBUU04 model~\cite{chen05}.}}
\label{lifig1}
\end{figure}

Using a parabolic approximation which has been verified by all
many-body theories to date, the equation of state (EOS) of isospin
asymmetric nuclear matter can be written as
\begin{equation}  \label{ieos}
E(\rho ,\delta )=E(\rho ,\delta =0)+E_{\rm sym}(\rho )\delta ^{2}
+\mathcal{O}(\delta^4),
\end{equation}
where $\delta\equiv(\rho_{n}-\rho _{p})/(\rho _{p}+\rho _{n})$ is the
isospin asymmetry and $E_{\rm sym}(\rho)$ is the density-dependent nuclear
symmetry energy. The latter is very important for understanding
many interesting astrophysical problems~\cite{lat01,steiner05}, the
structure of radioactive nuclei~\cite{brown,stone}, and the dynamics of
heavy-ion reactions~\cite{ireview98,ibook01,dan02,ditoro}. Unfortunately,
$E_{\rm sym}(\rho)$ is poorly known with theoretical predictions diverging
widely at both low and high densities as shown in the left panel of
Fig.~\ref{lifig1} for some of the most widely used microscopic many-body
theories~\cite{diep}.  The theoretical uncertainties are largely due to
a lack of knowledge about the isospin dependence of nuclear effective
interactions and the limitations of existing many-body techniques.
On the other hand, heavy-ion reactions, especially those
induced by radioactive beams, provide a unique opportunity to pin
down the density dependence of nuclear symmetry energy in terrestrial
laboratories. Indeed, significant progress has recently been made
both experimentally and theoretically in determining the symmetry energy 
at subnormal densities~\cite{chen05,betty04,steiner,lichen05,chenkl}. 
Future high energy radioactive beams will further allow one to determine the
symmetry energy at supranormal densities.

To extract information about the density dependence of the nuclear symmetry
energy from nuclear reactions induced by neutron-rich nuclei, reliable
reaction models that take into account the isospin degree of freedom
are needed. One such model that has been used extensively is the 
isospin- and momentum-dependent IBUU04 transport model~\cite{lidas03}. 
In this model, the single-nucleon potential, which is one of the most 
important inputs to all transport models, has the following density- 
and momentum-dependent form~\cite{das03}
\begin{eqnarray}
&&U(\rho,\delta,{\bf p},\tau ,x)=A_{u}(x)\frac{\rho _{\tau ^{\prime }}}{%
\rho _{0}}+A_{l}(x)\frac{\rho _{\tau }}{\rho _{0}}  \nonumber  \label{mdi}
+B\left(\frac{\rho }{\rho _{0}}\right)^{\sigma }(1-x\delta ^{2})\nonumber\\
&&-8\tau x\frac{B}{\sigma +1}\frac{\rho ^{\sigma -1}}{\rho _{0}^{\sigma }}
\delta \rho _{\tau^{\prime }}+\sum_{t=\tau,\tau^\prime}
\frac{2C_{\tau ,t}}{\rho _{0}}\int d^{3}{\bf p}^{\prime }\frac{%
f_{t}({\bf r},{\bf p}^{\prime })}{1+({\bf p}-{\bf p}^{\prime
})^{2}/\Lambda ^{2}}.
\label{gogny}
\end{eqnarray}
where $x$, $A_u(x)$, $A_{\ell}(x)$, $B$, $C_{\tau,\tau^{\prime}}$, $\sigma$, 
and $\Lambda$ are all parameters. The corresponding momentum-dependent
interaction (MDI) gives an incompressibility of $K_0=211$ MeV for
symmetric nuclear matter at saturation density. The parameter $x$ is 
introduced to mimic predictions on $E_{\rm sym}(\rho ) $ by microscopic 
and/or phenomenological many-body theories~\cite{lidas03}. Shown in
the right panel of Fig.  \ref{lifig1} is the density dependence of the
symmetry energy for $x=-2$, $-1$, $0$ and $1$. The momentum dependence
of the single-nucleon potential, given by the last term of
Eq.~(\ref{gogny}), leads to reduced effective masses for nucleons and
consequently also the in-medium nucleon-nucleon (NN) cross
sections~\cite{lichen05,fr}. In the IBUU04 transport model, both
experimental free-space and in-medium NN cross sections can be used.

\subsection{Constraining the density dependence of the nuclear symmetry
energy by isospin diffusion}

\begin{figure}[th]
\includegraphics[width=6.5cm,height=5.5cm]{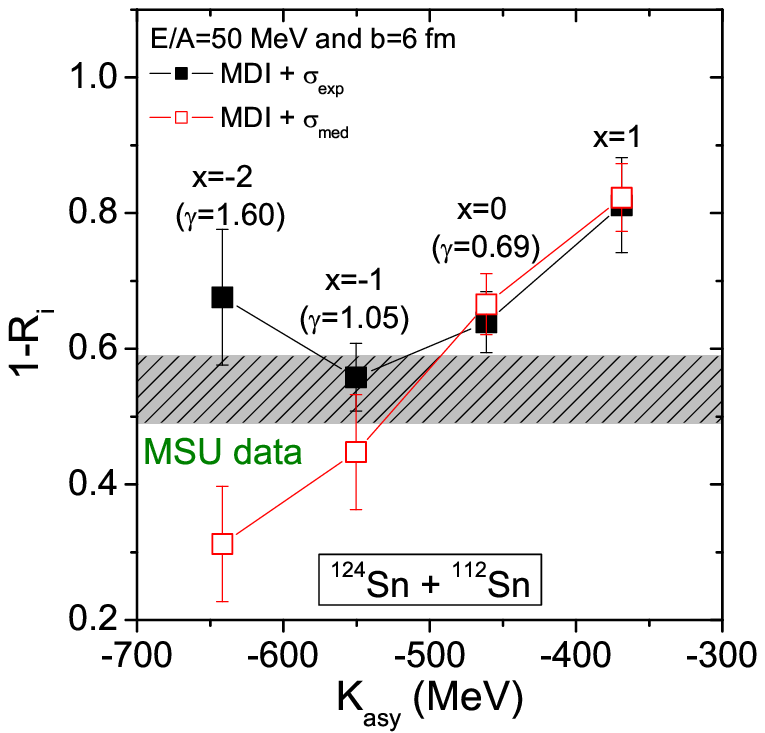}
\includegraphics[width=6cm,height=5.5cm]{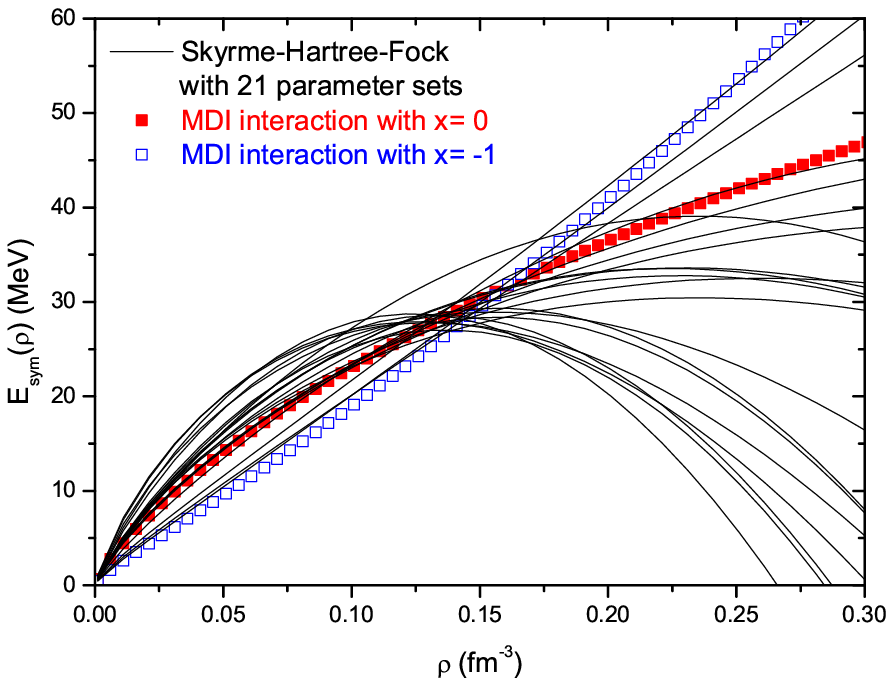}
\caption{{\protect\small Left panel: the degree of isospin diffusion as a
function of $K_{\rm asy}(\protect\rho _{0})$ for free (filled squares)
and in-medium (open squares) nucleon-nucleon cross sections~\cite{lichen05}.
Right panel: symmetry energies obtained from 21 sets of Skyrme 
interactions and the MDI interaction with $x=-1$ and $x=0$~\cite{chenkl}.}}
\label{lifig2}
\end{figure}

Tsang {\it et al.}~\cite{betty04} recently studied the degree of
isospin diffusion in the reaction $^{124}$Sn + $^{112}$Sn by measuring~
\cite{gsi}
\begin{equation}
{\rm R}_{\rm i}=\frac{2{\rm X}_{^{124}\mathrm{Sn}+^{112}\mathrm{Sn}}-
{\rm X}_{^{124}\mathrm{Sn}+^{124}\mathrm{Sn}}-{\rm X}_{^{112}\mathrm{Sn}
+^{112}\mathrm{Sn}}}{{\rm X}_{^{124}\mathrm{Sn}+^{124}\mathrm{Sn}}-
{\rm X}_{^{112}\mathrm{Sn}+^{112}\mathrm{Sn}}}
\label{Ri}
\end{equation}%
where ${\rm X}$ is the average isospin asymmetry $\left\langle \delta
\right\rangle $ of the $^{124}$Sn-like residue. Shown in the left
panel of Fig. \ref{lifig2} is a comparison of the strength of isospin 
diffusion $1-{\rm R}_{\rm i}$ obtained with free (filled squares) and 
in-medium (open squares) NN cross sections as a function of the 
asymmetric part of the isobaric incompressibility of nuclear matter 
at saturation density $\rho _{0}$~\cite{chen05}, i.e.,
\begin{equation}
K_{\rm asy}(\rho _{0})\equiv 9\rho _{0}^{2}\left( d^{2}E_{\rm sym}/
d\rho ^{2}\right)_{\rho _{0}}-18\rho _{0}\left( dE_{\rm sym}/d\rho
\right) _{\rho _{0}}.
\end{equation}
It is interesting to note that with in-medium NN cross sections the
strength of isospin diffusion $1-{\rm R}_{\rm i}$ decreases monotonically 
with decreasing value of $x$ (i.e., increasing stiffness). With free-space
NN cross sections, there appears, however, a minimum at around $x=-1$.
The value of $K_{\rm asy}(\rho_{0})$ determined from the use of in-medium
NN cross sections in the IBUU04 model is about $-500\pm 50$ MeV. 
This value is more precise than previous estimates~\cite{shlomo93}
and thus represents a significant advance in resolving the major 
remaining uncertainty about the incompressibility of nuclear 
matter at normal density~\cite{pie04,colo04}. Also shown in Fig. 
\ref{lifig2} are the values for $\gamma$ used in fitting the symmetry 
energy around $\rho_0$ using the expression 
$E_{\rm sym}(\rho )=32(\rho /\rho_{0})^{\gamma }$.  
From results obtained with the in-medium NN cross sections, the 
$\gamma$ parameter is constrained to be between $0.7$ and $1.1$.

\section{Constraining nuclear effective interactions}

The above constraints on the symmetry energy limit the nuclear
effective interactions in nuclear matter. This can be seen by
comparing them with the symmetry energies obtained from Skyrme 
effective interactions~\cite{chenkl}.  The right panel of 
Fig. \ref{lifig2} displays the density dependence of 
$E_{\rm sym}(\rho )$ for $21$ sets of Skyrme interaction parameters. 
Most of these effective interactions are not compatible with that extracted 
from the isospin diffusion data, given by the MDI interaction with 
$x$ parameter between $-1$ and 0, except the SIV, SV, G$_\sigma$, 
and G$_R$ Skyrme effective interactions.

\section{Constraining the slope of the symmetry energy}

\begin{figure}[th]
\includegraphics[width=5.4cm,height=6cm,angle=-90]{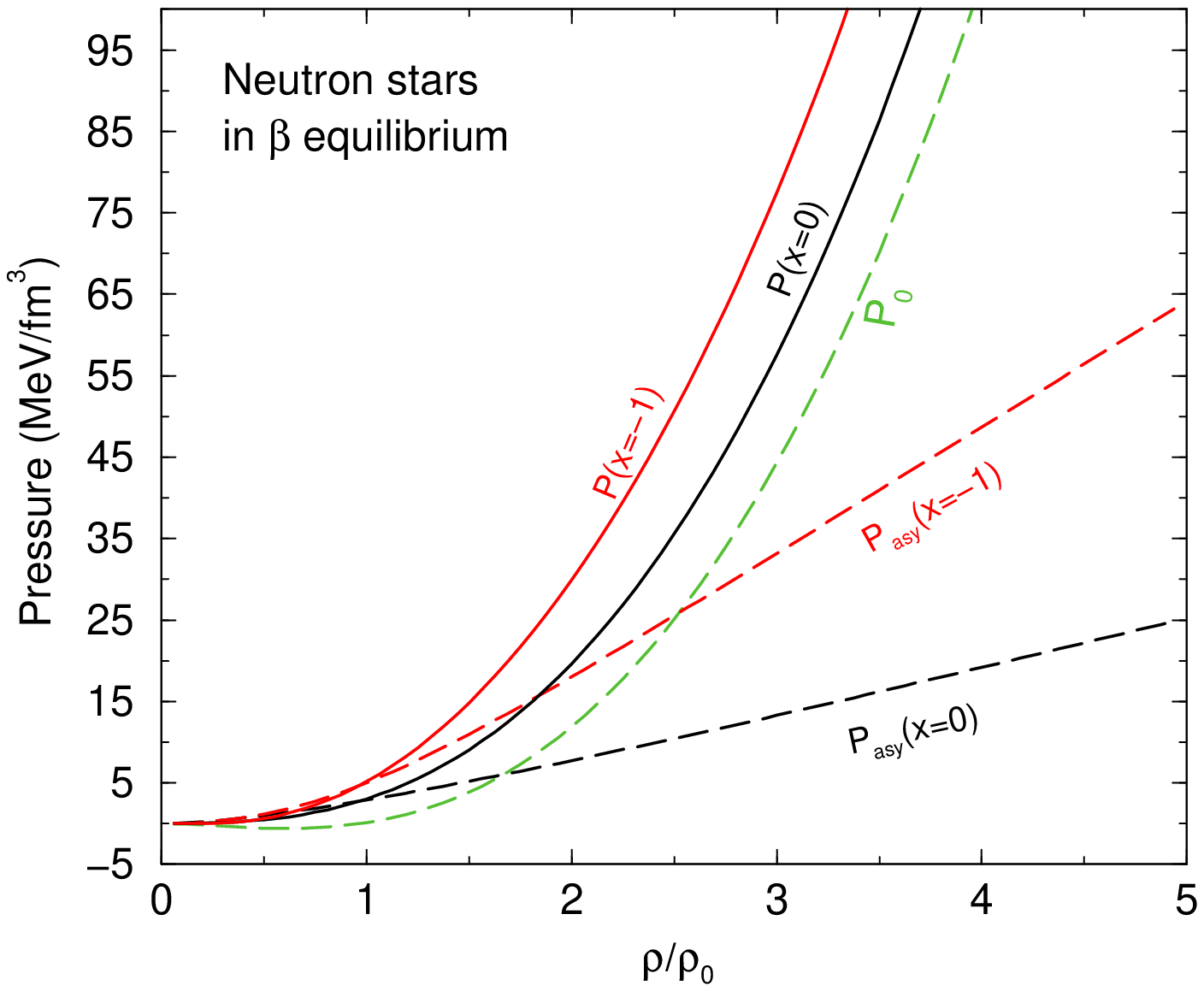}
\includegraphics[width=5.2cm,height=6cm,angle=-90]{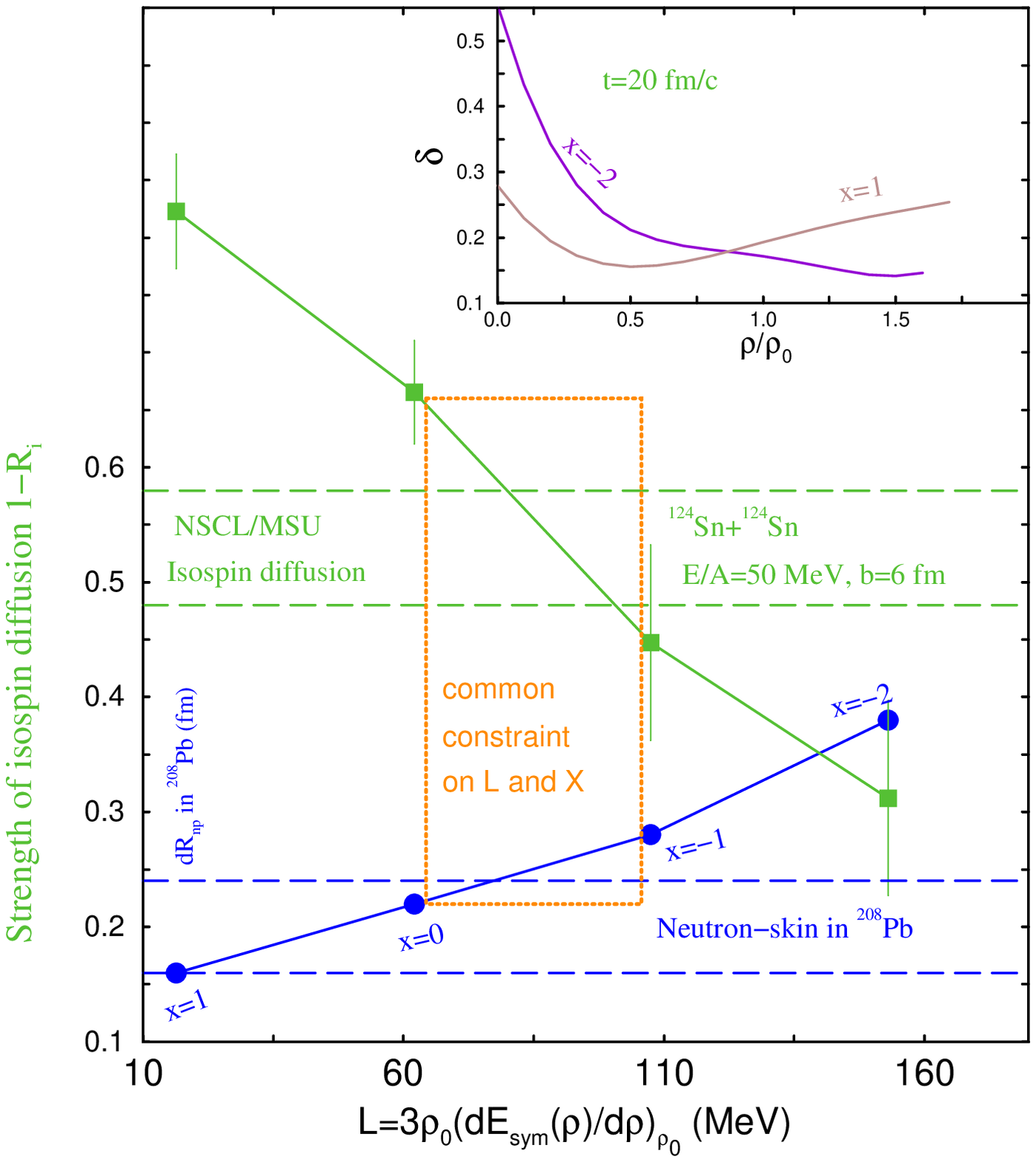}
\caption{{\protect\small Left panel: symmetric ($P_0$), asymmetric
($P_{\rm asy}$), and total pressure in neutron stars at 
$\beta$-equilibrium using the MDI interaction with $x=0$ and $x=-1$. 
Right panel: strength of isospin diffusion in $^{124}$Sn$+^{112}$Sn 
reactions and size of neutron skin in $^{208}$Pb as functions of the 
slope of the symmetry energy. The inset is the correlation between 
isospin asymmetry and density of matters at the instant of 20 
fm/c in the considered reaction~\cite{listeiner}.}}
\label{lifig3}
\end{figure}

While most properties of neutron stars depend on both the  
isospin-independent and -dependent parts of the nuclear equation of 
state, their radii are primarily determined by the isospin asymmetric 
pressure that is proportional to the slope of the symmetry energy
$E_{\rm sym}^{\prime}(\rho)$~\cite{lat01}.  For the simplest
case of a neutron-proton-electron ($npe$) matter in neutron stars
at $\beta$ equilibrium, the pressure is given by
\begin{eqnarray}\label{pre}
P(\rho,\delta)&=&P_0(\rho)+P_{\rm asy}(\rho,\delta)=
\rho^2\left(\frac{\partial E}{\partial \rho}
\right)_{\delta}+\frac{1}{4}\rho_e\mu_e\nonumber\\
&=&\rho^2\left[E'(\rho,\delta=0)+E'_{\rm sym}(\rho)\delta^2\right]
+\frac{1}{2}\delta(1-\delta)\rho E_{\rm sym}(\rho)
\end{eqnarray}
where the electron density is $\rho_e=\frac{1}{2}(1-\delta)\rho$ and
the chemical potential is $\mu_e=\mu_n-\mu_p=4\delta E_{\rm sym}(\rho)$.
The equilibrium value of $\delta$ is determined by the chemical equilibrium
and charge neutrality conditions, i.e., $\delta=1-2x_p$ with
\begin{equation}
x_p\approx 0.048 \left[E_{\rm sym}(\rho)/E_{\rm sym}(\rho_0)\right]^3
(\rho/\rho_0)(1-2x_p)^3.
\end{equation}
Because of the large $\delta$ value in neutron stars, the electron
degenerate pressure is small. Moreover, the isospin symmetric
contribution to the pressure is also very small around normal nuclear 
matter density as $E'(\rho_0,\delta=0)=0$. Shown in the left panel of
Fig. \ref{lifig3} are the isospin symmetric ($P_0$) and asymmetric
($P_{asy}$) as well as the total pressure in neutron stars at
$\beta$-equilibrium calculated from Eq.(\ref{pre}) using the MDI interaction
with $x=0$ and $x=-1$. It is seen that up to about $2.5\rho_0$ for 
$x=-1$ and about $1.5\rho_0$ for $x=0$ the total pressure is dominated 
by the isospin asymmetric contribution. Because neutron star radii are 
determined by the pressure at moderate densities where the proton 
content of matter is small, they are very sensitive to the slope of 
the symmetry energy near and just above $\rho_0$. In particular, a 
stiffer symmetry energy is expected to lead to a larger neutron 
star radius. 

For isospin diffusion in heavy-ion reactions, it is a re-distribution
of isospin asymmetries that is initially carried by the colliding nuclei.
The degree and rate of this process depend on the relative
pressures of neutrons and protons, namely the slope of
$E_{\rm sym}(\rho)$. With a stiffer $E_{\rm sym}(\rho)$,
it is harder for neutrons and protons to mix, leading thus to a
smaller/slower isospin diffusion.  Because of isospin asymmetric 
pressure, dilute neutron-rich clouds surrounding a more symmetric
dense region are dynamically generated in heavy-ion reactions through 
isospin diffusion as illustrated in the inset of the right panel of 
Fig. \ref{lifig3}, where the correlation between isospin asymmetry and 
density of matters at the instant of 20 fm/c in $^{124}$Sn$+^{112}$Sn 
reactions is shown. A similar effect is seen in neutron-rich nuclei, 
where the isospin asymmetric pressure pushes extra neutrons further 
out of an isospin symmetric core of near normal density, leading to 
an increasing of the sizes of neutron-skins in neutron-rich nuclei 
with increasing slope of the symmetry energy~\cite{steiner05,brown,chuck}.

From above discussions, one expects that the radii of neutron stars, 
the degree of isospin diffusion in heavy-ion collisions, and
the sizes of neutron-skins in heavy nuclei are all correlated through 
the same underlying nuclear symmetry energy. This is demonstrated
in the right panel of Fig. \ref{lifig3}, where the strength of the 
isospin diffusion $1-{\rm R_i}$, calculated with the IBUU04 model with 
in-medium NN cross sections, and the thickness of neutron skin 
$dR_{np}$ in $^{208}$Pb, calculated using the Skyrme Hartree-Fock 
with interaction parameters adjusted to give an EOS which is similar 
to the effective interaction used in the IBUU04 model~\cite{steiner}, 
are examined simultaneously as functions of the slope parameter 
$L\equiv3\rho _{0}(\partial E_{\rm sym}/\partial \rho)_{\rho_0}$
of the symmetry energy. It is seen that $1-{\rm R_i}$ decreases while 
$dR_{np}$ increases with increasing $L$ as expected. Taken the fiducial 
value $dR_{np}=0.2\pm 0.04$ fm, that is measured and supported
strongly by several recent calculations~\cite{steiner05},
and the NSCL/MSU data $1-{\rm R_i}=0.525\pm0.05$~\cite{betty04},
the $L$ parameter is constrained in a common range between
62.1 MeV ($x=0$) and 107.4 MeV ($x=-1$)~\cite{listeiner}.

\section{Constraining the radii and cooling mechanisms of neutron stars}

\begin{figure}[th]
\includegraphics[width=6cm,height=5.5cm]{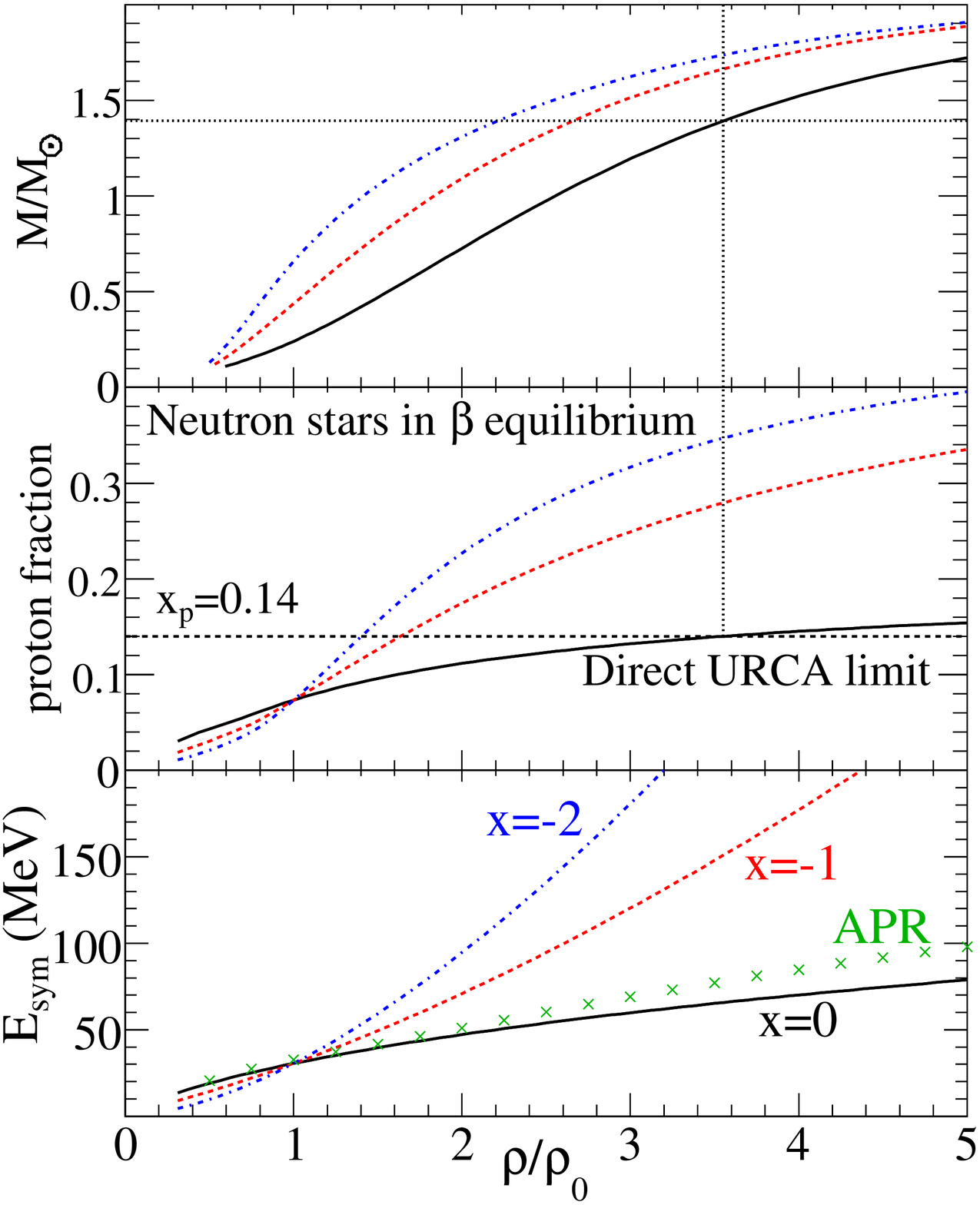}
\includegraphics[width=6cm,height=5.7cm]{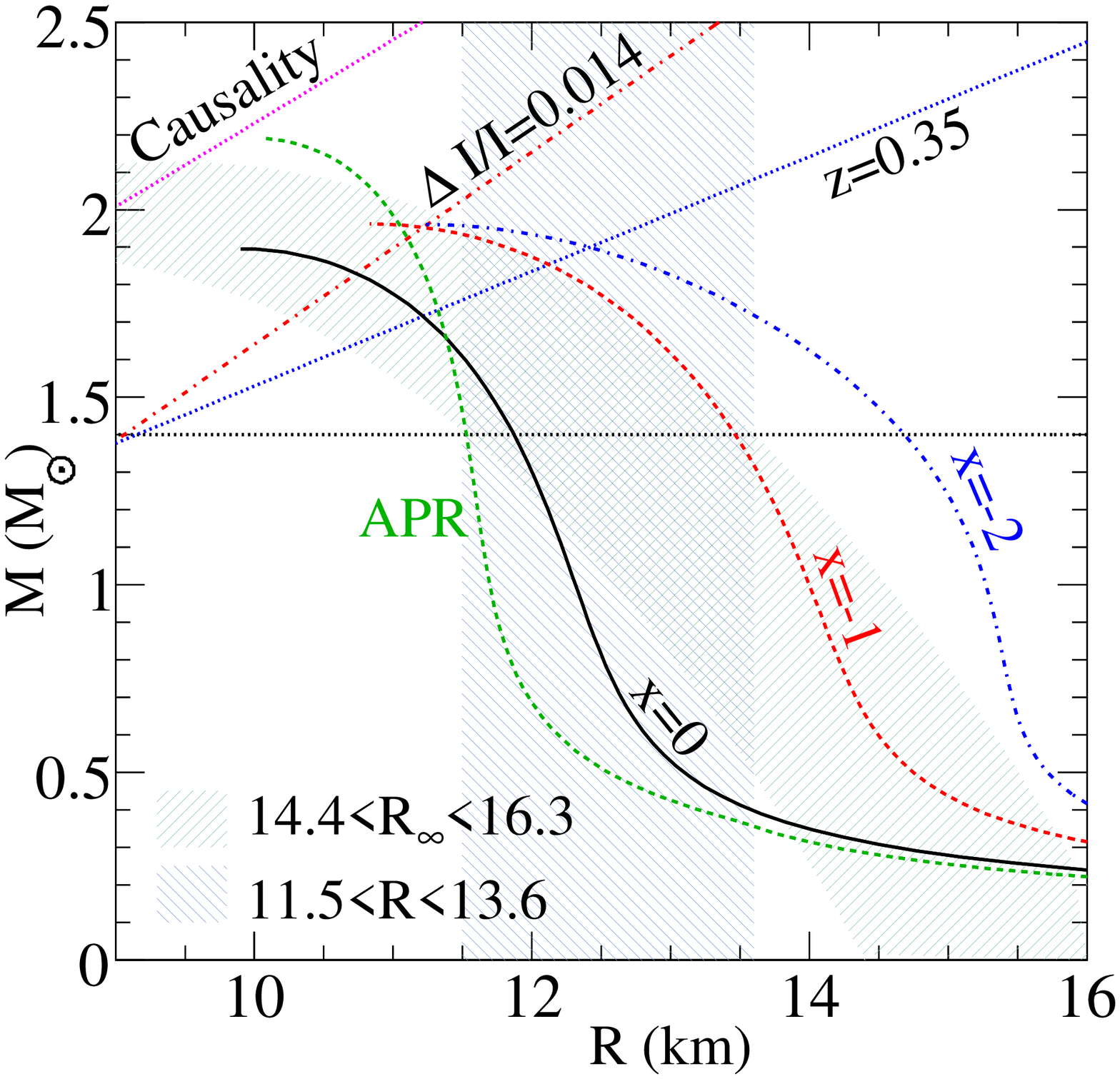}
\caption{{\protect\small Left panel: mass and proton fraction of
neutron star and the symmetry energy as functions of density.
Right panel: correlations in neutron star masses and radii~\cite{listeiner}.
All results are for spherically-symmetric, non-rotating, non-magnetized 
neutron stars consisting of $npe\mu$ matter at zero temperature.}}
\label{lifig4}
\end{figure}

The dependence of some of the basic properties of neutron stars on the
nuclear EOS is shown in the left panel of Fig. \ref{lifig4}. 
The top and middle panels give, respectively, the mass of neutron star 
and its proton fraction $x_p$, calculated from the MDI interaction 
with $x=0, -1$, and $-2$, as functions of its central density.
The symmetry energies obtained from these interactions are shown in 
the lower panel together with that from the AV18+$\delta v$+UIX$^{*}$ 
interaction of Akmal {\it et al.} (APR)~\cite{apr}. It is interesting 
to see that up to about $5\rho_0$ the symmetry energy predicted by 
APR agrees very well with that from the MDI interaction with $x=0$ 
and is thus within the range of symmetry energies constrained by both 
isospin diffusion and $^{208}$Pb neutron-skin thickness data. 
Because of the lack of experimental information on the EOS of pure 
neutron matter, the APR prediction has been widely used as a 
{\it standard} in calibrating other microscopic calculations. 

For $x_p$ below 0.14~\cite{lat01}, the direct URCA process for fast 
cooling of proto-neutron stars does not proceed because energy and 
momentum conservation cannot be simultaneously satisfied. For EOSs 
from the MDI interaction with $x=-1$ and $x=-2$, the condition for 
direct URCA process is fulfilled for nearly all neutron stars
above 1 \Msun. For the EOS from $x=0$, the minimum density for the
direct URCA process is indicated by the vertical dotted line, and the
corresponding minimum neutron star mass is indicated by the horizontal
dotted line.  Neutron stars with masses above 1.39 \Msun~are thus
expected to have a central density above the threshold for the
direct URCA process.

The relations between neutron star masses and radii for above EOSs 
are given in the right panel of Fig. \ref{lifig4}.  Also given 
are the constraints due to causality as well as the mass-radius 
relations from estimates of the crustal fraction of the moment of 
inertia ($\Delta I/I=0.014$) in the Vela pulsar~\cite{link} and 
from the redshift measurement from Ref.~\cite{cottam}. Allowed 
equations of state should lie to the right of the causality line 
and also cross the other two lines. The hatched regions are inferred 
limits on the radius and the radiation radius (the value of the 
radius which observed by an observer at infinity) defined as 
$R_{\infty}=R/\sqrt{1-2GM/Rc^2}$ for a 1.4~\Msun~neutron star.  
It is seen that the symmetry energy affects strongly the radius of 
a neutron star but only slightly its maximum mass~\cite{lat01,pra88}. 
These analyses have led to the conclusion that only radii between 
11.5 and 13.6 km (or radiation radii between 14.4 and 16.3 km) are 
consistent with the EOSs from the MDI interaction with $x=0$ and 
$x=-1$ and thus with the laboratory data on isospin diffusion and 
neutron skin thickness of heavy nuclei~\cite{listeiner}.

The observational determination of the neutron star radius from the
measured spectral fluxes relies on a numerical model of the neutron
star atmosphere and uses  as inputs the composition of the atmosphere, a
measurement of the distance, the column density of x-ray absorbing
material, and the surface gravitational redshift. Since many of
these quantities are difficult to measure, only a paucity of 
radius measurements are available.  Current estimates obtained from 
recent x-ray observations give a wide range of results. It 
is therefore very useful to compare our results with recent 
Chandra/XMM-Newton observations. Assuming a mass of 1.4 \Msun, the
inferred radiation radius $R_{\infty}$ (in km) is $13.5\pm
2.1$~\cite{bob02} or $13.6\pm 0.3$~\cite{gendre02a} for the neutron
star in $\omega$ Cen, $12.8\pm 0.4$ in M13~\cite{gendre02b},
$14.5^{+1.6}_{-1.4}$ for X7 in 47 Tuc~\cite{ryb05}, and
$14.5^{+6.9}_{-3.8}$ in M28~\cite{becker}. Except the
neutron star in M13 that has a slightly smaller radius, all others
fall into our constraints of 14.4 km $<R_{\infty}< 16.3$ km within the
observational error bars that are larger than the above range in most cases.

\section{Summary}
The EOS of neutron-rich matter is fundamentally important for both nuclear
physics and astrophysics. Nuclear reactions induced by neutron-rich nuclei
provide a great opportunity to constrain the EOS of neutron-rich matter.
The recent analysis of isospin diffusion data in heavy-ion collisions within
the IBUU04 transport model allowed us to constrain the density dependence
of the nuclear symmetry energy in a narrow range. Together with experimental 
information on the neutron-skin thickness of $^{208}$Pb, this has led to
stringent constraints on not only the nuclear effective interactions but also 
the cooling mechanisms and radii of neutron stars.

\bigskip

This Work work was supported in part by the US National Science 
Foundation under Grant No. PHYS-0354572 and PHYS-0456890, and 
the NASA-Arkansas Space Grants Consortium award ASU15154 (BAL); 
National Natural Science Foundation of China under Grant No. 10105008 
and 10575071 (LWC); the US National Science Foundation under Grant
No. PHY-0457265 and the Welch Foundation under Grant No. A-1358 (CMK); 
and the US Department of Energy under grant No. DOE/W-7405-ENG-36 (AWS).

\end{document}